\documentclass[a4paper]{article}
\usepackage{graphicx}    % For graph
\usepackage[T1]{fontenc} % Modern font encoding
\usepackage{float}       % For creating charts, graphs and schemes
\usepackage{helvet}      % Helvetica font for sans serif
\usepackage{mathptmx}    % Times font ("Word-like")

\usepackage{fullpage}
\usepackage{rsc}
\usepackage{authblk}
\usepackage[usenames,dvipsnames]{xcolor}
\begin{document}
%%%TITLE, AUTHORS AND ABSTRACT%%%

\noindent
\LARGE{\textbf{A genomic characterization of metallic nanoparticles}} 
\\%Article title goes here instead of the text "This is the title"
\noindent\large{Kevin Rossi $\dagger$, Gian Giacomo Asara $\dagger$, and Francesca Baletto $\dagger$}
\footnote{Corresponding author: francesca.baletto@kcl.ac.uk}  
\\%Author names go here instead of "Full name", etc.
{\textit{$\dagger$ Physics Department, King's College London, Strand, WC2R 2LS, UK}}

\noindent
\begin{abstract}{
\noindent
With a focus on platinum nanoparticles of different sizes (diameter of 1-9~nm) and shapes, we sequence their \textit{geometrical genome} by recording the relative occurrence of all the non equivalent active site,
classified according to the number of neighbours in their first and second coordination shell.
The occurrence of sites is morphology and size dependent, with significant changes in the relative occurrence up to 9~nm. 
Our geometrical genome sequencing approach is immediately transferable to address the effects of the morphological polydispersivity in size-selected samples and the influence of temperature, including ionic vibrations and thermal activated processes. 
The proposed geometrical genome forecasts an enhancement of the catalytic reduction of molecular oxygen on stellated and anisotropic platinum twinned nanoparticles, with their shortest axes of $\sim$2~nm, and an irreversible disruption of the Pt nanocatalyst's structure above 1000~K.}
\end{abstract}

%%%END OF TITLE, AUTHORS AND ABSTRACT%%%
%%%END OF FOOTNOTES%%%

%%%%
%Nomenclature
%  Each active site and adsorption mode is characterised by a different GCN value; 
% the GCN fingerprint identifies the list of all GCN values for a certain geometry
% the relative frequency/occurrence/rate of a certain GCN value gives information on the cluster's activity
% the GCN fingerprint can be associated to the cluster selectivity, as intermediate products can be stable only on some sites
% the GCN fingerprint and the relative abundance of each GCN peak/value can define a geometrical genome for a cluster

%%%MAIN TEXT%%%%
\noindent
\section{Introduction}
One of the main challenges that our society faces is the need to find green and renewable sources of energy. This requires the extensive use of nanocatalysts to improve green chemistry reactions such as the slow oxygen reduction in fuel cells. In this effort, it is vital to move from the current description to a prediction at the atomic level of catalytic properties, leading to novel designing rules.\cite{Catlow2016}  
Ranging across chemistry, materials science, mathematics, biology and engineering, catalysis is one of the most interdisciplinary of all branches of science. The "Ertl's revolution" started in the Seventies, when surface science techniques allowed the understanding of catalytic processes at the molecular level\cite{Ertl2007}, found one of his maximum potentials thanks to the advent of nanocatalysts (NCs) almost thirty years ago. NCs combine the highly efficiency and selectivity of heterogeneous catalysts with the easy recovery/recycling property of the homogeneous catalysts, with the spread of two new families: colloidal catalysts referring to coated nanoparticles in solution, and metallic nanoparticles (mNPs) relating mostly to the heterogeneous catalysis domain. By itself nanotechnology is one of the most exciting and recent developments of materials science. While metallic nanoparticles (mNPs) represent the building blocks of heterogeneous nanocatalysts, nanoparticles are quite attractive as their applications span over a much wider and cross-disciplinary horizon.
Strong structural effects with frequent shape fluctuations among the variety of geometries -fluxionality at the nanoscale \cite{Alexandrova2017}- are due simply to the lack of any translational symmetry constraint. This translates in strained chemical bonds, electron delocalization, which may vary with NP size and shape, electronic bands, which may not be fully formed. In short, mNPs constitute a new state of matter. The significant advances in microscopy and spectroscopy techniques allow the structure of mNPs to be resolved at the atomic level, monitor its evolution and offer the possibility to perform in-situ studies of the chemistry and structure of mNPs in a reactive environment \cite{Palmer2016,Wang2012}. Although fully engaged with a "nano-mania", the beauty and amazing features of NPs are still poorly understood in the sense that we still ignore how to predict and exploit the size and morphology effects to achieve control of a certain property, such as their catalytic activity. The nanocatalyst's performance can be understood by determining the reaction mechanisms that lead to the different intermediates and products. These mechanisms often consist of many steps, for which a full mechanistic study is excessively demanding by standard computational tools \cite{Ertl2008}. The use of descriptors, however, can accelerate the evaluation of the catalytic performance by analysing a given feature of the active sites and reactants \cite{Abild-Pedersen2016}. Descriptors are indeed theoretical derived quantities which establish a correlation between structure and a chemophysical property, e.g. the adsorption energy, thus playing a central role for the design of new catalysts with better properties. Here we propose a full characterisation of various mNP geometrical motifs elucidating its application towards catalysis.  

Numerical guidelines and receipts would avoid a costly and ineffective trial-and-error procedure pursuing the synthesis of samples with a well defined but suboptimal architecture (size, shape, chemical composition and ordering).
Modern design schemes should also account for the intrinsic geometrical heterogeneity of the sample and the variety of adsorption sites proper of each structural motif \cite{Davis2015}. The latter influences the activity and the selectivity of each cluster \cite{Seh2017}, while the former can affect the overall performance of the device. The aim of this work is to show that the prediction of nanocatalyst's life-time and reactivity can be mapped into an atomistic understanding of how dynamical structural changes affect the geometrical descriptor(s), a radical change in heterogeneous catalysis where usually the catalyst is considered static. In other words, we aim to provide an accurate geometrical analysis under operando conditions which forecasts the catalytic behaviour of individual nanoparticles as well as polydiverse samples.

We predict stellated decahedra as the optimal shape of free Pt-clusters for catalysing the reduction of molecular oxygen, one of the bottleneck for the widespread use of fuel cells in the automotive and energy industry. Our predictions are in agreement with recent experimental results that point out the enhanced catalytic activity, structural stability and durability of complex twinned nanoarchitectures compared to standard Pt/C catalyst \cite{Huang2017, Alia2017, Bian2015}. 
%Stable noble metal twinned and stellated nanoparticles are synthesised using simple one-pot procedure.\cite{Liu2014}Small lattice mismatch and controlled procedures allows an epitaxial growth of the catalytic active Pt skin which preserves the underlying twinned symmetry\cite{Bian2015} producing a very active core-shell catalyst. The characteristic enhanced activity can be also found in other low dimensional twinned structures like nanorods and nanowires. }\cite{Huang2009, Huang2017, Alia2017}
 
 \section{Decoding a geometrical genome for metallic nanoparticles}
\begin{figure*}[t!]
\begin{center}
\includegraphics[width=12cm]{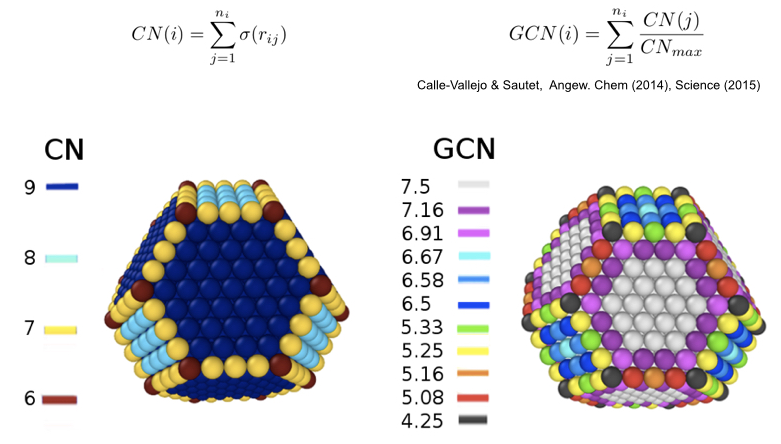}
\end{center}
\caption{{\bf Choice of GD}:  Coloured mapping and genome of atop adsorption site onto a regular truncated octahedron as it derived from using the coordination number, namely the number of nearest neighbour atoms, (left) and the generalised coordination number, which includes the effect of the second neighbour shell as it is the sum of the coordination of all the nearest neighbour of a site $i$ weighted with the referencevalue in the bulk (right). }
\label{Fig:GCN_vs_CN}
\end{figure*}
Our analysis starts with a detailed classification of all the non equivalent adsorption sites (NEAS) enabling a rationalisation of the catalytic activity as a function of mNPs' architecture -size, morphology. Let us first discuss how to choose a geometrical descriptor (GD) at the core of this classification method. This turns to be fundamental as the GD choice has to satisfy different requirements, to correctly distinguish the NEAS of a nanoparticle and encode a clear dependency on the cluster's size and shape. At the same time, it should be able to distinguish among adsorption modes. At least the chosen GD should map of atop/bridge active sites in clusters presenting different size and morphologies. To stressing the importance of the GD we have compared the classification for atop adsorption sites arisen from the coordination number (CN) and the generalised coordination number (GCN), as introduced by Calle-Vallejo \cite{Calle-Vallejo2014}, which in some extents average over the second shell of second neighbours. The comparison is reported in Figure \ref{Fig:GCN_vs_CN} with a net enhacement of details using the GCN instead of CN, as the former is able to distiguish sites close to edges and vertexes that unlikely behaves as low-Miller indexes sites.

We would like to note that the GCN seems to be powerful for the characterisation of mNP and it turns to be suitable as descriptor for the adsorption energy of small molecules O$_2$, OH, CO, both on monometallic \cite{Calle-Vallejo2014, Asara2016, PazBorbon2017} and bimetallic nanoparticles, but it could fail in providing a good reference for adsorption of longer molecule as ethanol, see Ref. \cite{Rigo2018}, as in those cases the relative position of the tail with respect to the nanoparticle must be considered. This features has not a trivial dependence on the coordination of the metallic anchor/site and hence cannot be easily mapped by the GCN. 

\begin{figure*}[t!]
\begin{center}
\includegraphics[width=16cm]{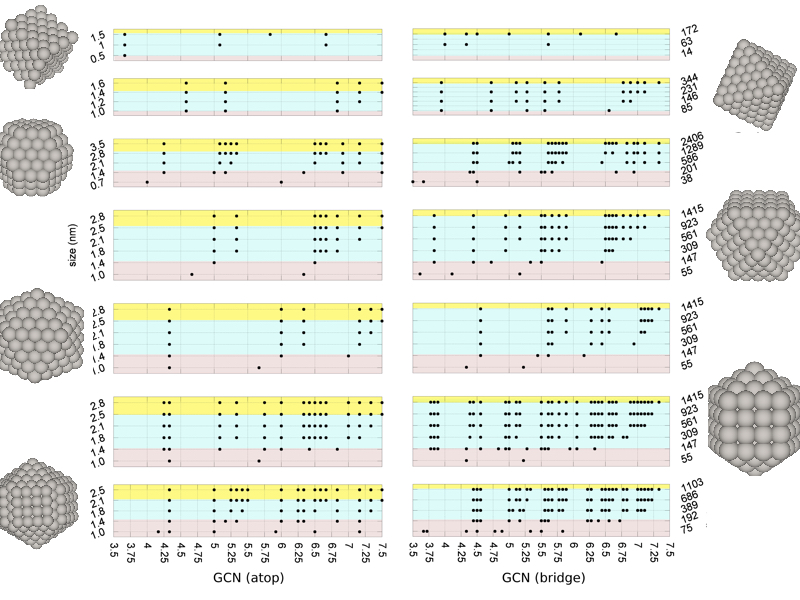}
\end{center}
\caption{{\bf GCN Fingerprint}: Top row: An array of atop (left) and bridge (right) GCN fingerprints for Marks decahedra at different sizes. Each NEAS are identified by their GCN value. At small size NEAS relative amount between high and low GCN values is fairly homogeneous but we notice the fingerprint is very mobile. When increasing the cluster size, the fingerprint presents fixed characteristic and it can be used to univocally identify a nanoaprticle.}
\label{Fig:finger_bridge}
\end{figure*}

The catologue of each type of NEAS and counting their number leads the sequencing of a \textit{geometrical genome} for each nanocluster motif. This catalogues the type and counts each non equivalent adsorption site present in a cluster, as depicted in the lower panel of Figure \ref{Fig:finger_bridge}.
 \begin{figure*}[t!]
 	\begin{center}
 \includegraphics[width=6.5cm]{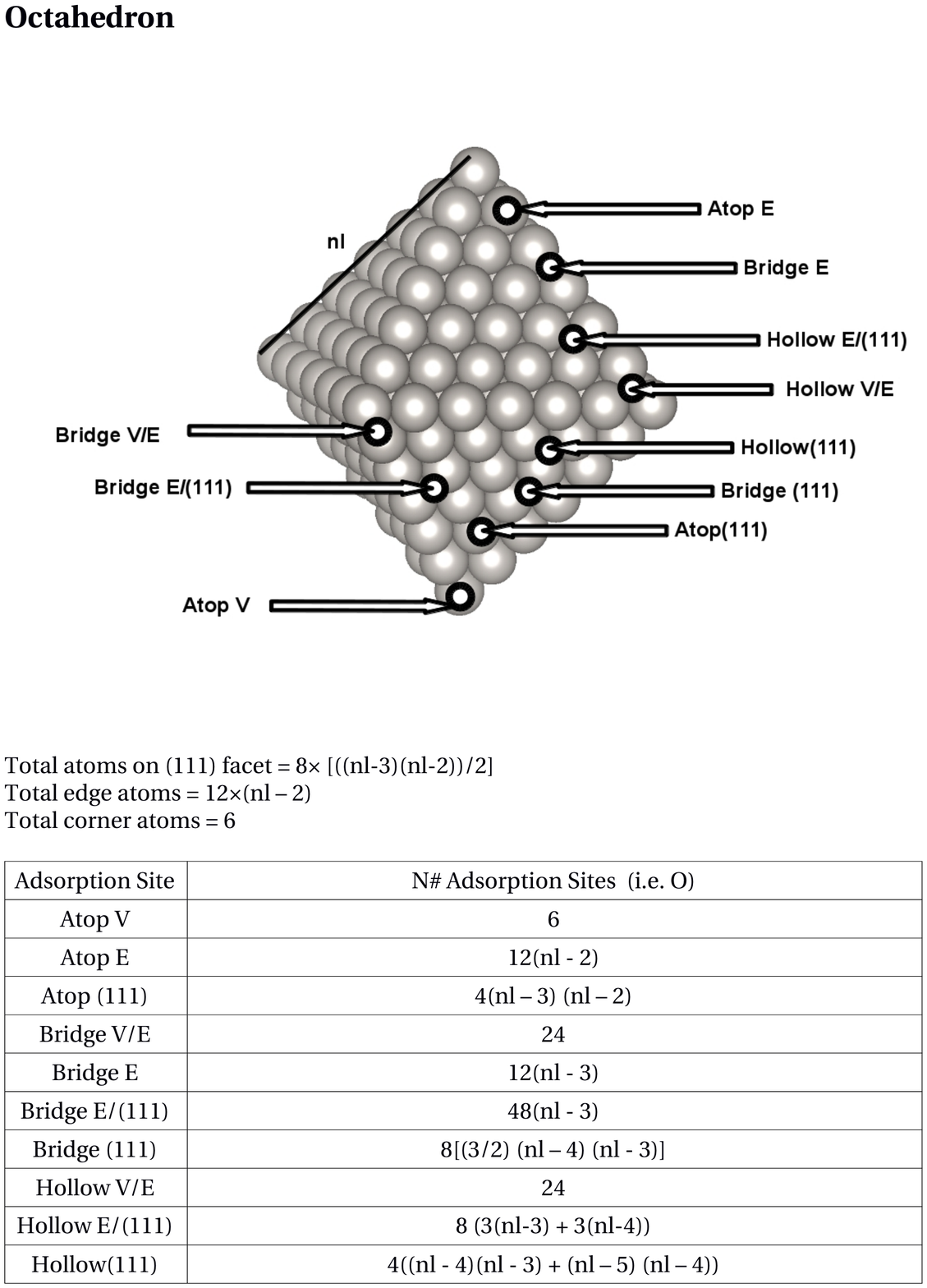}
 \includegraphics[width=6.5cm]{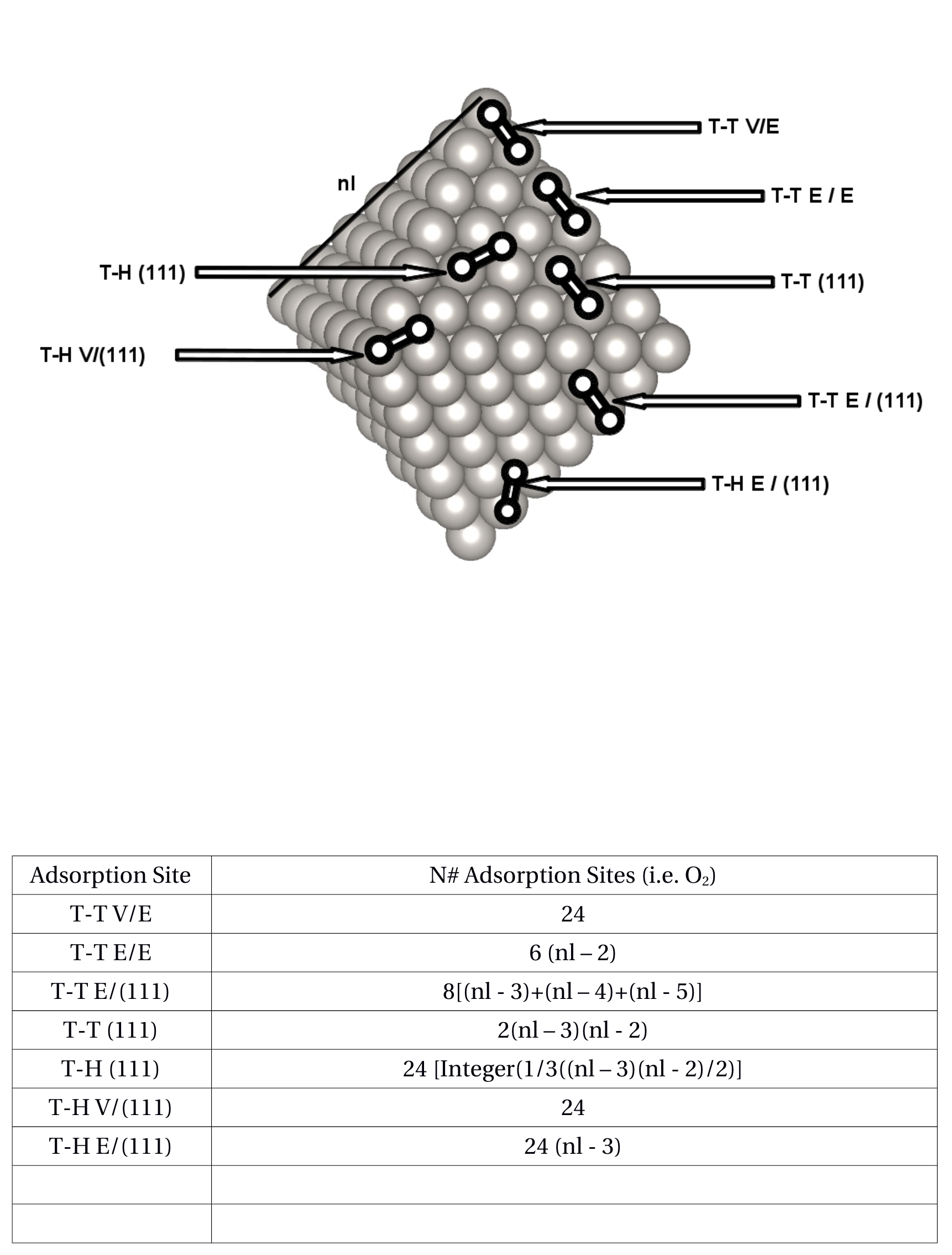}
 \end{center}
 \caption{{\bf Counting of NEAS for a Oh}: A visual representation of the counting procedure of NEAS for a Oh. Both atop and bridge cases are depicted on the left right panel, respectively.}
 \label{Fig:counting}
 \end{figure*}
Even at the cost of being pedantic in Figure \ref{Fig:counting}, we do report how NEAS can be cassified and then counted for the paradigmatic case of an octahedron for both atop and bridge sites. The classification is shown to follow a geoemtrical and physical meaning. We show how the counting/occurence can be performed as a function of the number of atoms along the edge. 

A systematic high trough-put sequencing of geometrical genomes for architectures spanning over a wide range of sizes and shapes and the identification of structure-property scaling relationship, able to rationalise adsorption energies and activation barriers in terms of the geometry of the active site. The geometrical genomes can be then used to screen and select case-studies NEAS where adsorption should be investigated explicitly by means of ab-initio calculations.
Successfully pursuing both enables to screen promising architectures and to propose rational design guidelines for advanced applications. still with an eye to the design of metallic nanocatalysts, we would like to mention the possibility to use this approach within a kinetic Monte Carlo framework \cite{Schmidt2018}.
With a focus on platinum nanoparticles, we sequenced the geometrical genome of seven different structural motifs: the cube (Cb), the octahedron (Oh), its regular truncation (rTo), the cubocathedron (Co), the icosahedron (Ih), and cuts of a pentagonal bipyramid, in the literature known as the Ino-Decahedra (IDh), and the Marks-Decahedra (MDh). Structures are depicted in Figure \ref{Fig:GCN_vs_CN}, while the main result of this investigation is reported in Figure \ref{Fig:finger_bridge}: where the size evolution of the fingerprint is shown for MDh and the geoemtrical genome of the eight motifs is drawn following a standard convention in genomics: on the left side of the circle the 'blueprint' values are reported on the right the shapes we want to classified. The width of each band represent the abudance of a given fingerprint. Below each shape the full genome is reported.
Far from be exhaustive of all archetypical motifs at the nanoscale, we believe that the ensemble of morphologies here considered provides a sufficient structural variety to test the effectiveness of our approach.

We observe a general pattern in the size dependent evolution of the list of NEAS, see Figure \ref{Fig:finger_bridge}, as distinguished by their GCN, present and characteristic of each structure. We will refer to this quantity as the GCN fingerprint of the cluster. Independently of the shape and adsorption mode, we identify three size-regions according to the diameter of the cluster, d$_{cl}$, highlighted in the plot in different colours:
\begin{itemize}
\item d$_{cl} <$ 1.5~nm, only few NEAS available, the GCN fingerprint is not only incomplete but also mobile, with several peaks displaced with respect to the ones at larger sizes.
\item 1.5 $<$ d$_{cl}$ $<$ 3.5~nm, the GCN fingerprint is not fully complete yet, because the facets are not large enough to accommodate all the different active sites, e.g. GCN > 7 are not shown yet.
\item d$_{cl} >$ 3.5~nm, all the considered shapes show a complete GCN fingerprint  which constitutes a distinctive geometrical attribute of that morphology.
\end{itemize}
The three size-regimes point out at which cluster size the electronic charge distribution at an active site is likely to be similar to its surface counterpart, if existing, and to predict the properties of sites in systems prohibitively expensive to be studied at ${\it ab-initio}$ level from the ones in smaller clusters.

Features in the GCN fingerprint can be rationalized in terms of the cluster morphology and size.
When delimited by a variety of facet, nanoparticles show a diversification of their NEAS with
the MDh here having the largest number due to the presence of both (111), (100), and re-entrant facets.
Motifs which share a given local environment, also present NEAS with the same GCN value.
Nanoparticles presenting facets with a larger (smaller) area to perimeter present earlier (later) a complete GCN fingerprint.
Size and shape interplay also drives the relative occurrence of NEAS.
Vertex and close-to vertex sites are the first to appear but their number is fixed, they are followed by the edge NEAS, and close-to-edge NEAS, which scale linearly with size, and finally by surface-like ones which present a quadratic dependence with respect to cluster nuclearity. A non trivial size dependence of the relative occurrence of each NEAS thus arises.
As shown in Figure \ref{Fig:finger_bridge} for d$_{cl}$ = 1.5~nm we observe an almost homogeneous relative occurrence of each NEAS, with active sites with GCN in the 4-6 range more frequent than others. At d$_{cl}=$3.5~nm low-Miller index surface sites are the more frequent, only in the case of the To and the Cb, yet also in this case, they total only for the 22\% and the 50\% of the active sites. Only at 5.5~nm all morphologies display surface sites as their most abundant, with their relative occurrence being at least twice with respect to the other sites for d$_{cl}$=7.5~nm.
Cu, Oh, and Ih present the smaller number of different NEAS being terminated only by either (100) or (111) facets.
Co, IDh, and To display a larger one because they are limited by both (100) and (111) facets.
For example the active sites with GCN(atop)=7.5, found in all the morphologies here considered apart Cu, lies on (111) facets. Similarly active sites with GCN(atop)=6.6, corresponding to sites on (100) surfaces are identified in all apart Ih, and Oh. Only concave structures present NEAS with GCN (atop) $>$ 7.5.%, where O adsorbates experiment optimal binding energy for ORR catalysis.

\begin{figure*}[t!]
\begin{center}
	\includegraphics[width=12cm]{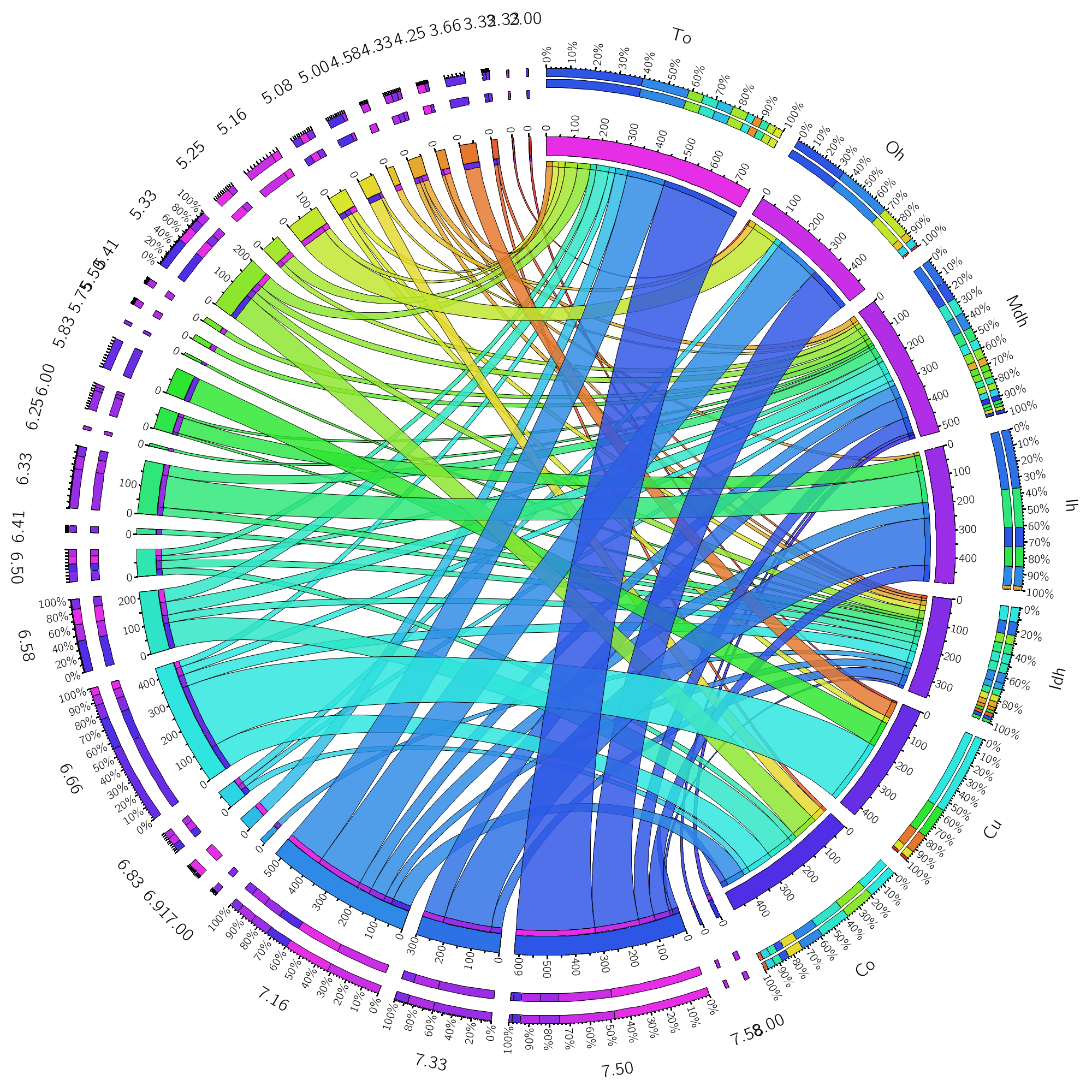}
\end{center}
\caption{{\bf GCN genome}:  Bottom row: Their collection leads to a fingerprint characteristic of the architectures at a sufficiently large nanoparticles once they have their unique fingerprint. The occurrence (here in \%) of each site-type -GCN value- is represented by the width of the band connecting shape and GCN value.}
\label{Fig:genome}
\end{figure*}

%%%%%%%%%%%%%%%%%%%%%%%%%%%%%%%%%%%%%%%%%%%%%
% FIG 3 a geometrical disperse sample & temperature: EXPERIMENTS
\begin{figure*}[t!]
	\begin{center}
\includegraphics[width=18cm]{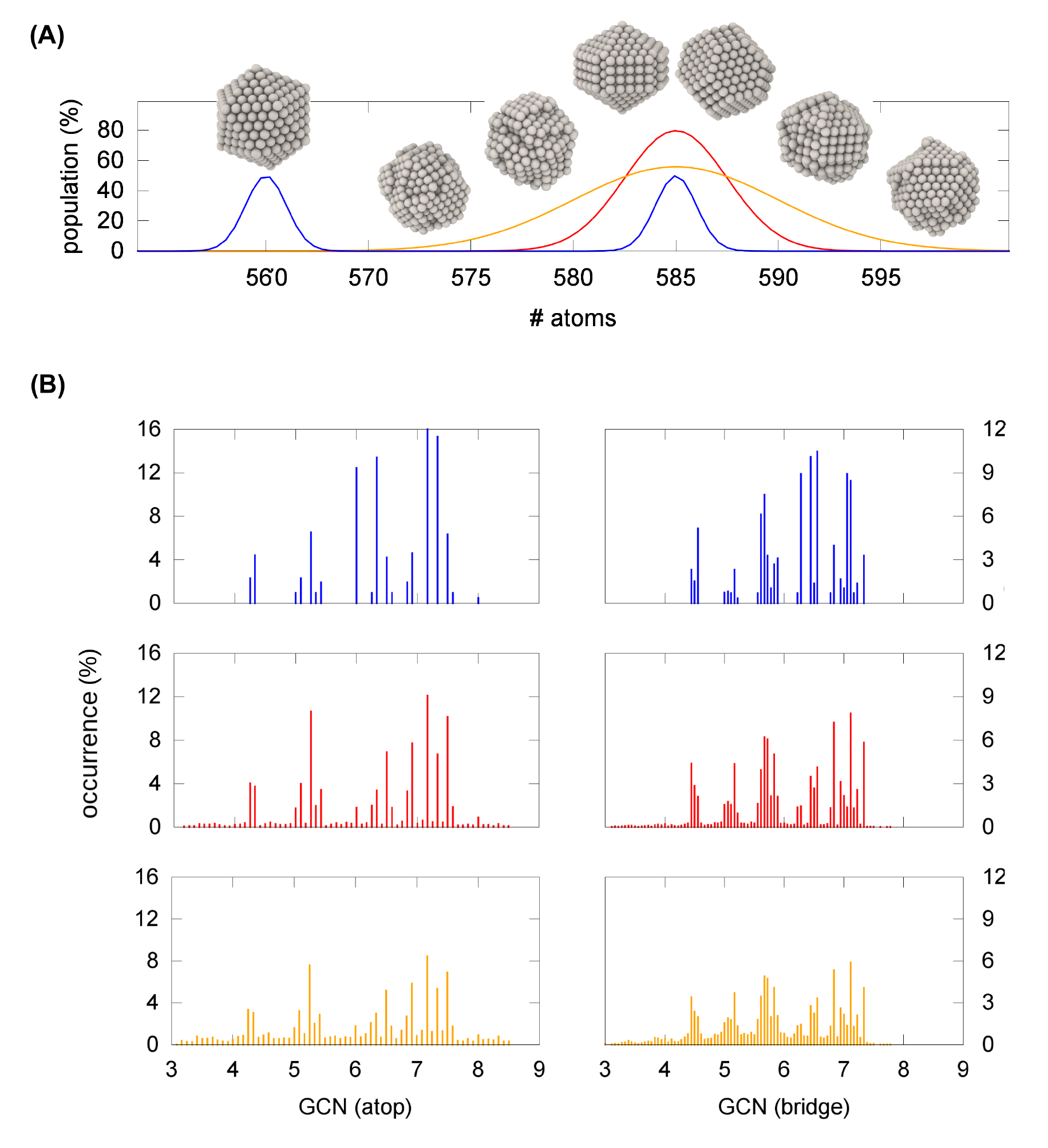}
\end{center}
\caption{{\bf GCN genome analysis of a geometrical diverse sample}: A visual representation of the three populations discussed in the text in terms of their representative clusters' distribution and shapes (A). The GCN genome of each population for atop and bridge adsorption modes (B). The more the sample loses its morphological identity, the broader its GCN genome becomes, presenting new values at both low and high GCN, and decreasing the occurrence of its most abundant NEAS.}
\label{Fig:gcn_sample}
\end{figure*}

\section{From one to many: monodisperse sample analysis}
To provide specific geometrical descriptor genome based design guidelines beyond the static and ideal motif picture, we propose two gedanken experiments: the first is devoted to reveal the effect of geometrical polydispersivity of a size-selected sample, the second aims to elucidate the role by ionic vibrations, surface diffusion and shape fluctuations at finite temperatures. The latter suggests the need of developing models which include the mobility of the surface atoms and the reconstruction of the adsorption sites. The former calls for a comprehensive investigation of the effect of morphological diversity in ensembles of clusters, and how it affects their catalytic properties.

To show how the GCN genome depends on the geometrical heterogeneity in a sample, we compare three populations of Pt-nanoparticles with 575$\pm$20 atoms, as in the top panel of Figure \ref{Fig:gcn_sample}.
In this size range, from energetic considerations based on the standard and modified Wulff construction\cite{Ringe2013c}, MDh of 585 atoms are expected to be formed.
On the other hand, at specific magic sizes Ih and To, of 561 and 586 atoms respectively, should be energetically competitive with the MDh and likely to appear in a size-selected sample for the chosen size range.
From Figure \ref{Fig:finger_bridge}, we note that Ih has still a negligible number of surface-like sites, To has already a 20-25\% of highly coordinated sites, and MDh allows for the adsorbtion of molecules in its re-entrances.
Population (i), in red, is described by a unimodal distribution peaked around 585 atoms with a standard deviation of 5 atoms.
It is composed of an 80\% equally split between MDh$_{585}$ and To$_{586}$, respectively, while the remaining 20\% are low energy minima at 580 and 590 atoms, as found by means of basin-hopping search\cite{Rossi2009} (see Section 2 for technical details).
For population (ii), in orange, we allow a standard deviation of 15 atoms with the 60\% of MDh$_{585}$ and To$_{586}$, another 30\% made of the putative global minima at 580 and 590 atoms, respectively, and the last 10\% of energetically favourable shapes of 595 and 575 atoms.
Population (iii), in blue, displays a bimodal distribution of closed-shell geometries centred at 561, where Ih should be energetically favourable, and at 585 and 586 atoms, where we find respectively a MDh and a rTo. The relative abundance of each structure is 50\% for the Ih and 25\% each for MDh and To.
When analysing the GCN genome of each population, we note that samples of closed-shell structures, as the blue population, are characterised by fewer NEAS, with overall higher GCN values. 
Introducing defected shapes, with a variety of re-entrances, steps, and islands, the GCN genome shows a broadened distribution, with tails towards both smaller GCN values as well as higher ($>$ 7.5)  due to the presence of concave adsorption sites.
Comparing the red and orange populations, we note that even a small change in the cluster size and shape distribution may significantly affect the population GCN genome, hence a non-negligible difference in their catalytic performance is expected.
When the size distribution and the morphological heterogeneity of the ensemble are broadened, the GCN values centred around 5 lessens from 25\% to 20\%, and the percentage between of the ones between 7 and 8 falls from 50\% to 30\%.
We can extrapolate the qualitative catalytic behaviour of a sample by looking at its ensemble GCN genome: a morphologically disperse nanocatalyst is likely to show a lower selectivity, with many side-reactions taking place.
An increase in the sample geometrical heterogeneity would produce a nanocatalyst with a broader range of applications while a properly shape selected nanocatalyst will be highly selective for a target application.
We remark that this experiment can be easily generalised to include other isomers, to weight the population accordingly to their thermodynamical stability or against their energetic stability in contact with an environment; to extend it to other size ranges, to be  compared with experimental data.

\section{Ionic mobility at the nanoscale}
\begin{figure*}[h!]
\begin{center}
\includegraphics[width=18cm]{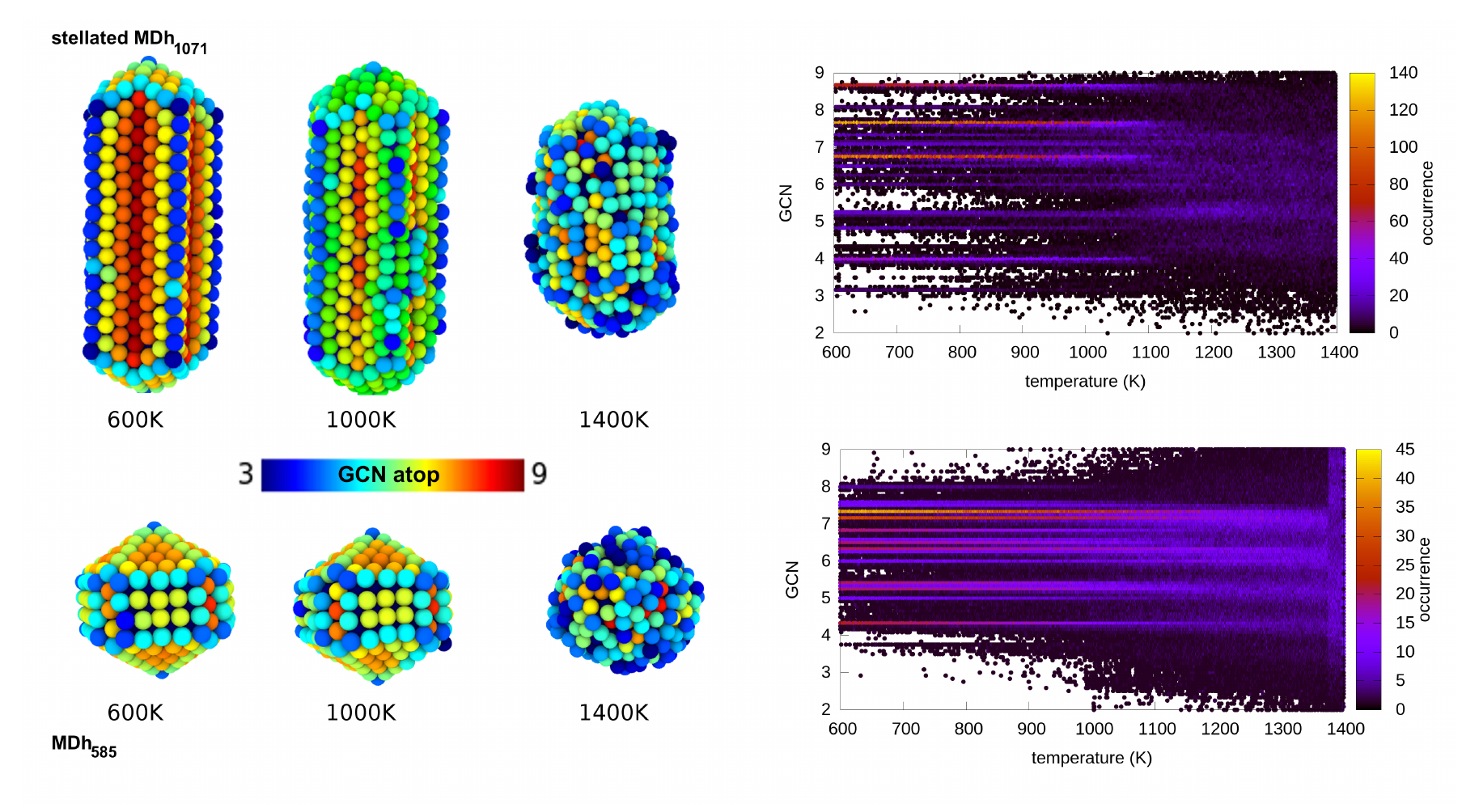}
\end{center}
\caption{ {\bf Effect of the thermal ionic motion on GCN genome}
Snapshots of the structures of the initial elongated stellated MDh$_{1071}$ ($m=1, n=15, p=4$) and MDh$_{585}$ ($m=5, n=4, p=2$) at 600~K,1000~K,100~K where atop active sites are coloured according to their GCN (A). 
Time/temperature evolution of atop GCN genome for elongated stellated MDh$_{1071}$ and MDh$_{585}$ during a melting procedure, data points are coloured according to the occurrence of NEAS presenting that particular atop GCN value (B).
The formation of defects is associated to the appearance of NEAS with GCN different from the ones of the closed sell structure.
Surface diffusion, happening between 1000-1300~K is highlighted by a smoothing of the GCN genome.
The melting transition corresponds to a continuum in the GCN fingerprint.}
\label{Fig:gcn_t}
\end{figure*}

The GCN genome can be also analysed during the numerical modelling of formation processes, such as growth, coalescence, and nucleation, where the effect of the environment, including solvent, ligands, and support can be encoded too. 
Continuous monitoring of this quantity can provide important hints for the rational synthesis of nanocatalysts specifically optimised for a target application. 
Here, as the first and somehow preliminary step towards the modelling of complex and realistic scenarios, we analyse the effect of the ionic vibrations, and thermally activated processes, such as surface diffusion and island formation, on the GCN genome of two MDh clusters.
We consider gas-phase monometallic Pt systems of 585 and 1071 atoms and investigate their structural evolution when subject to a slow heating from 600~K to 1400~K by means of classical Molecular Dynamics, using the software \texttt{LoDiS}, where the interatomic potential is dderived within the second moment approximation of the tight-binding\cite{Baletto2002}. Here we are using a iterative temperature algorithm for mimicking the melting of the nanoparticles, as described in \cite{Rossi2017, Rossi2017a, Rossi2018, Pavan2015}. The code is available upon request to the authors.

In Figure \ref{Fig:gcn_t} left side, we plot the temperature (time) evolution of the GCN genome for atop NEAS, whose occurrence is highlighted by different colours.
The GCN genome is essentially fixed up to low temperatures as 600~K. Vibrations of surface atoms lying on (111)/(111) edges lead to a first broadening of the GCN genome in the 600-1000~K range and the appearance of various other NEAS with GCN $<$ 5.
Between 1000-1300~K, the vibrational motion of (111) surface atoms becomes relevant and the characteristic value for surface-like sites fade.
We stress that no major structural rearrangements are observed in these temperature ranges although surface diffusion may take place, Figure \ref{Fig:gcn_t} right side.
The original GCN genome is replaced by a more uniform distribution of GCN values with slightly more populated sites (~10-15\%) around 7, 6, 5 and 4 while the peak close to 3 disappears. The onset of melting is evident slightly below 1300~K, where all sites become almost equiprobable and the GCN genome looses any specific feature, Figure \ref{Fig:gcn_t} right side. We expect a change of the catalytic activity against the external temperature for bare clusters.

\section{Tailored Pt nanoparticles for ORR}
To illustrate and clarify how the proposed approach can lead to a rational nanocatalyst design, we focus on the oxygen reduction reaction (ORR) on Pt nanoparticles.
ORR is an important electrochemical process which occurs at the fuel cell cathode, with the chemisorption of molecular oxygen and involving many intermediates (e.g. O, OOH, OH, H$_2$O$_2$).
It is characterised by a very low reaction rate which prevents a wider use of low temperature fuel cells as a reliable power source. 
State-of-the-art ${\it ab-initio}$ calculations demonstrated that the generalised coordination number is a powerful descriptor, in predicting the adsorption trends of oxygenated molecules on active sites in gas phase Pt nanoparticles \cite{Calle-Vallejo2014, Calle-Vallejo2015, Calle-Vallejo2015a, Calle-Vallejo2017} and PtNi alloyed clusters supported on oxide \cite{Asara2016}: there exist a linear relation between the binding energy of ORR intermediates at the adsorption site  and the correspondent GCN value, i.e. the bond weakens increasing the coordination.
Because of the well known relationship between the adsorption energy of OH (and atomic O) with the overall ORR activity of Pt catalysts \cite{Norskov2009a}, the established linear dependence between adsorption energy and active site GCN can be used to successfully forecast the catalytic activity of each NEAS in the nanoparticle with a maximum associated to the ones with GCN between 7.33-8.5.\cite{Calle-Vallejo2017} 
The nanoparticle genome sequencing according to the generalized coordination may also enable a deeper understanding of the trends found in the specific activity, defined as the catalyst turn over frequency normalised with respect to the catalyst active area.
Within the GCN characterization framework, active sites at the vertexes of the clusters and in their first neighbourhood are fixed in number, proper edge sites and the facet ones in their first coordination shell scale linearly with the size of the cluster and only proper faces NEAS scale quadratically, as expected from a geometrical argument.

\begin{figure*}[t!]
	\begin{center}
\includegraphics[width=12cm]{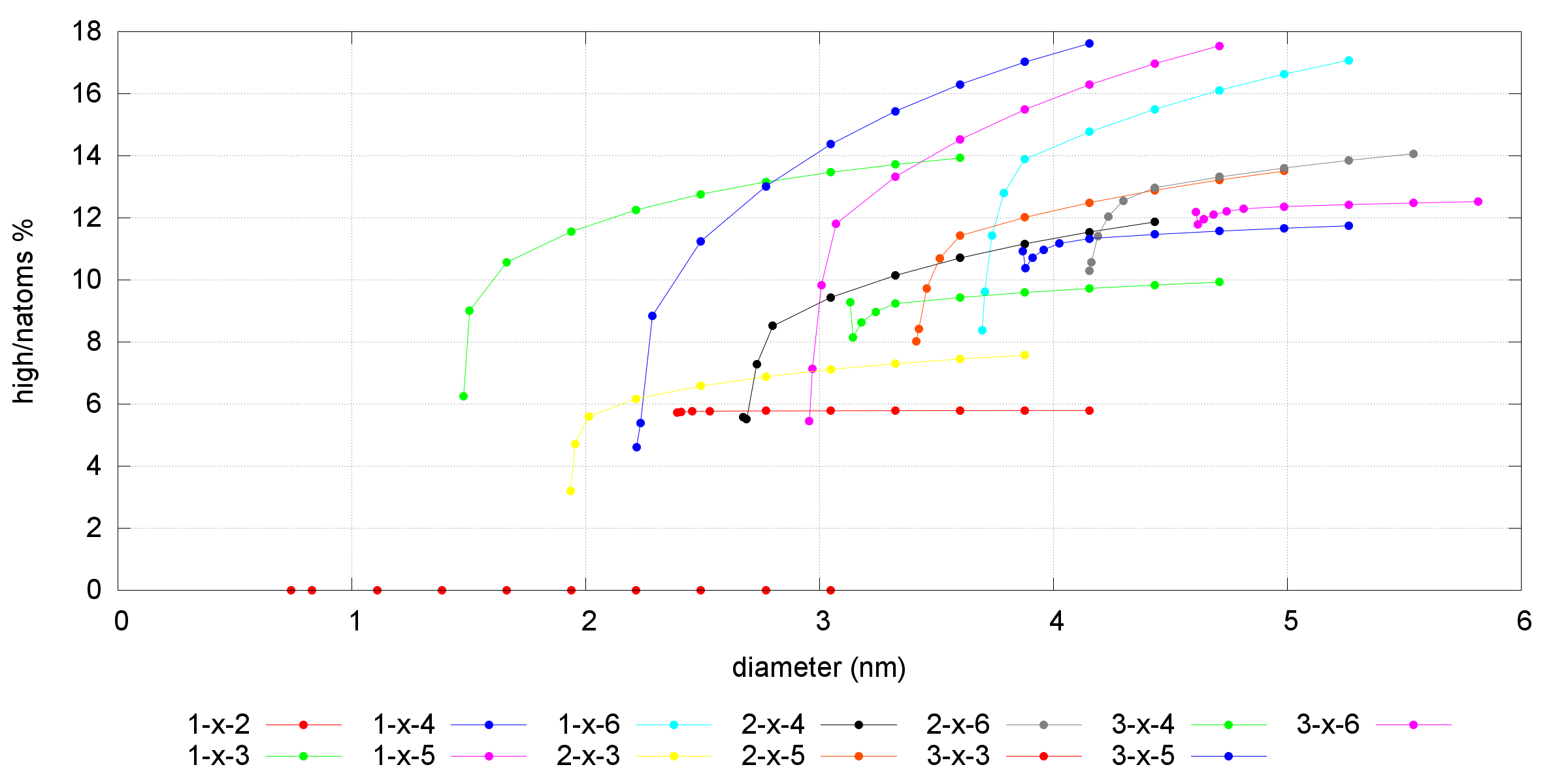}
\end{center}
\caption{{\bf Selection of Stellated decahedra}: A visual representation of how the density of highly GCN sites varies against several decahedra, with $m$ between 1 and 3; $p$ has values in the 2-6 range while $n$ is the independent variable in the plot.}
\label{Fig:gcn_MDH}
\end{figure*}

The estimate of size dependent NEAS properties by means of our GCN genome sequencing is in good agreement with accurate electronic structure calculations recently reported by N{\o}rskov's group: the adsorption properties of atomic O and molecular CO over (111) and (211)-like sites onto a platinum Co with d$_{cl} >$ 1.5 ~nm (147 atoms) resemble their extended surface counterpart.\cite{Li2013}
Looking at the size dependence of the genome, we can predict the mass activity for ORR on Pt nanoparticles to show significant drop for cluster below 2~nm in diameter. This can be easily explained in terms of the lack of active sites characterised by a GCN between 7.33-8.5, see \ref{Fig:finger_bridge}.
Furthermore, low generalised coordinated sites are likely to undergo non-reversible oxidation thus de-activating the catalyst. Although the consensus is not universal, it is commonly found that the mass activity for clusters peaks between 2.2 and 3~nm.\cite{Shao2011, Li2013, Kleis2011}
At this size a synergic variety of NEAS is found, with a balanced relative occurrence of each. 
Increasing the  size of the nanoparticle, the specific activity is known to gradually fade to the one of the extended low-Miller index surfaces.
This consideration can be inferred from the GCN genome of convex structures, where the number of active sites on Low-Miller index facets growths more rapidly than the edge and vertex sites.
For anisotropic and concave structures where the re-entrances length is comparable to the one of the short axes, the ratio between convex active sites (high coordination, GCN (atop) > 7.5) to low-Miller index facets ones (GCN(111)=$7.5$, GCN(100)=$6.6$) is at least $\sim$ 0.5 also for large sizes. Hence the activity is predicted to be enhanced respect to low Miller index surfaces. 
%Similarly, golden cuboactahedra show finite-sizes effects only for d$_{cl} <$ 2.5 ~nm (561 atoms), above this size adsorption energies and electronic charge distribution on (111) and (211)-like sites agree with their counterpart on the corresponding infinite surface.\cite{Tritsaris2011} 
Following the same counting routine shown for Oh, Fig. \ref{Fig:counting}, we have investigated how we can cut a decahedron -or more generally a pentagonal bi-pyramid- in order to create concavities that seems promising for ORR. The numerous available truncations of a Dh can described by three integers $m, n$ and $p$ identifying the number of atoms along the five-fold axis, along the (100) edge perpendicular to it, and along the twin planes - often called Marks-reentrance- respectively.
By systematically exploring the $m, n, p$ parameters space, we observe that the relative occurrence of NEAS with GCN (atop) $>$ 7.5 (respect to the other surface sites and the total number of atoms) is optimized for anisotropic structures elongated along their five-fold axis, with a shorter diameter of 2~nm and a longer of at least 3~nm. In Figure \ref{Fig:gcn_MDH} we report a graphic comparison between the different series of Marks decahedra, highlighting the density with respect to the number of total atoms in the NP, of highly GCN sites -the most promising for ORR in relationship with the available scaling realtionship.

The geometrical genome can help putting together the pieces of the puzzle and shed light on the controversy about which structural feature are crucial to enhance Pt NPs ORR activity: (i) stellated nanostructures demonstrated to be extremely active (\cite{Liu2014}), and their high activity is attributed to the high density of vertex, edges low coordinated sites and surface area; (ii) it has been demonstrated active sites can be created \emph{ad~hoc} with single atom removal from a pristine (111) surface (\cite{Calle-Vallejo2015}), as well as at the interface between to coalesced Pt NPs (\cite{Calle-Vallejo2017}), highlighting the importance of \emph{increasing} the coordination to increase activity; (iii) the geometrical genome identifies the best candidates in concave "stellated" clusters because of the presence of high coordinated sites on the re-entrances, which are predicted to be very active (\cite{Calle-Vallejo2017}). We hope our work will inspire and motivate new experiments that can unequivocally determine the role of concave sites on the ORR activity. We claim that the high activity measured on stellated nanoarchitectures has been erroneously attributed to the presence of low coordinated sites when it was determined by the abundance of highly coordinated counterparts (with GCN>7.5). Within the structures explicitly considered in this work, such high GCN value can be found only for sites inside the characteristic re-entrance of Marks decahedra.

\section{Conclusions}
In summary we show that practical guidelines for the rational design of Pt nanoparticle for ORR can be extracted by sequencing a geometrical genome based on the CN of adsorption sites, as we did for nanoparticles of various shapes and diameter up to 9~nm.
We focus on common experimentally synthesized architectures for Pt systems and identify three activity-size regimes according to the variety and relative abundance of the active sites.
This characterization offers a novel perspective to rationalize the trends for oxygen reduction reaction in terms of the relative abundance and type of their NEAS. The relative stability of the intermediates involved in the process depends upon the coordination of the adsorption site. The predominant reaction mechanisms, dissociative or associative,  is likely to be linked to and influenced by the relative percentage of low and high coordination sites such that a thermally driven increment in the population of NEAS with low (GCN < 4) or high (GCN > 7) coordination will force the reaction towards one or the other mechanism \cite{Asara2016, Jia2014}; further, even at temperature where surface diffusion and structural rearrangements are rare, we show, for the first time, that the effect of ionic vibrations on the genome of the metallic nanoparticle may be significant. We suggest a sample rich in twinned Pt-nanoparticles with pronounced re-entrance, very small, if not absent, (100) facets, possibly elongated along the five-fold axis, with the longest axis of at least 3~nm and a width of 2~nm: anisotropic stellated Marks decahedra maximise the number of highly coordinated sites per metallic load, hence being optimal candidates for ORR.
To reduce the negative effects associated to ionic vibrations and thermally activate processes, which determine a net broadening of the GCN genome of gas-phase platinum nanoparticles with the suggested shape and size, the catalyst working temperature should be kept well below 1000~K. An excessive geometrical diversity of the sample might play a detrimental effect, especially in terms of a lower selectivity.

We believe that this elegant and effective approach provides new directions to nanocatalysts' engineering, promotes the search for geometrical descriptors able to encode the active sites dependence on the local environment and their effect on the adsorbate stability. It reveals the importance of addressing the impact of thermal activated processes on the structure of the nanoparticle, and delivers practical guidelines for the rational design of more active and selective nanocatalysts.
By systematically analysing the GCN genome of several architectures with different sizes and morphologies, we forecast truncated stellated anisotropic decahedra as optimal nanocatalysts for ORR. 

\section*{Acknowledgments}
GGA and FB thank the "Towards an Understanding of Catalysis on Nanoalloys" (TOUCAN) EPSRC Critical Mass Grant (No. EP/J010812/1) as does KR (Grant Reference ER/M506357/1). All the authors thank further the financial supported offered by the Royal Society (No. RG 120207). GGA is grateful to the postdoctoral scheme offered by the NMS Faculty.
\section*{Author contributions}
The work started from an original idea of FB. KR and GGA wrote the original code for the calculus of the generalized coordination number, FB and KR are the main developers of \texttt{LoDiS} and all the authors coded specific tools to generate geometrical structures. KR and GGA performed all the calculations. All the authors analysed the data and write the manuscript.

%It is likely selectivity for a certain reaction is enhanced only at specific NEAS, thus a catalyst polydisperse samples enriched with nanoparticles of sizes and shapes with a highly occurrence of such specific active sites will be highly desirable.

%%%%\section*{Methodology}

%%%REFERENCES%%%
%\bibliography{neas} %You need to replace "rsc" on this line with the name of your .bib file
%\bibliographystyle{rsc} %the RSC's .bst file
\providecommand*{\mcitethebibliography}{\thebibliography}
\csname @ifundefined\endcsname{endmcitethebibliography}
{\let\endmcitethebibliography\endthebibliography}{}

\end{document}